\documentclass[10pt,A4paper,conference]{IEEEtran}
\usepackage{amsthm}
\usepackage{amsmath}
\usepackage{amsfonts}
\usepackage{graphicx}
\usepackage{latexsym}
\usepackage{amssymb}
\usepackage{stmaryrd}
\usepackage{comment}
\usepackage{amscd}
\usepackage[small]{caption}
\usepackage[dvipsnames]{xcolor}

\newcommand{\sbinom}[2]{\left[ \begin{array}{c} #1 \\ #2 \end{array} \right] }

\newcommand{\field}[1]{\mathbb{#1}}

\newcommand{\Z}{\field{Z}}

\newcommand{\cA}{{\cal A}}

\newcommand{\cC}{{\cal C}}

\newcommand{\sG}{\script{G}}

\newcommand{\sP}{\script{P}}

\newcommand{\bfa}{{\boldsymbol a}}
\newcommand{\bfb}{{\boldsymbol b}}
\newcommand{\bfc}{{\boldsymbol c}}

\newcommand{\bft}{{\boldsymbol t}}

\newcommand{\bfx}{{\boldsymbol x}}
\newcommand{\bfy}{{\boldsymbol y}}
\newcommand{\bfz}{{\boldsymbol z}}

\DeclareMathOperator*{\argmax}{arg\,max}
\DeclareMathOperator*{\argmin}{arg\,min}

\newcommand{\abs}[1]{\left|#1\right|}

\DeclareMathAlphabet{\mathbfsl}{OT1}{cmr}{bx}{it}
\newcommand{\uuu}{\kern-1pt\mathbfsl{u}\kern-0.5pt}
\newcommand{\vvv}{\kern-1pt\mathbfsl{v}\kern-0.5pt}

\newcommand{\wt}[1]{ \text{wt} ({#1}) }

\newcommand{\myboxplus}{\kern1pt\mbox{\small$\boxplus$}}

\makeatletter \DeclareRobustCommand{\sbinom}{\genfrac[]\z@{}}
\makeatother
\newcommand{\G}[2]{\sbinom{{#1}\kern-1pt}{{#2}\kern-1pt}}
\newcommand{\Gq}[2]{\sbinom{{#1}\kern-0.25pt}{{#2}\kern-0.25pt}}

\newcommand{\Ps}{\smash{{\sP\kern-2.0pt}_q\kern-0.5pt(n)}}
\newcommand{\sPs}{\smash{{\sP\kern-1.5pt}_q(n)}}
\newcommand{\Ptwo}{\smash{{\sP\kern-2.0pt}_2\kern-0.5pt(n)}}
\newcommand{\Ptwom}{\smash{{\sP\kern-2.0pt}_2\kern-0.5pt(m)}}
\newcommand{\Ptwonm}{\smash{{\sP\kern-2.0pt}_2\kern-0.5pt(n+m)}}
\newcommand{\Ptwoa}{\smash{{\sP\kern-2.0pt}_2\kern-0.5pt(1)}}
\newcommand{\Ptwob}{\smash{{\sP\kern-2.0pt}_2\kern-0.5pt(2)}}
\newcommand{\Ptwoc}{\smash{{\sP\kern-2.0pt}_2\kern-0.5pt(3)}}
\newcommand{\Ptwod}{\smash{{\sP\kern-2.0pt}_2\kern-0.5pt(4)}}
\newcommand{\Ptwoe}{\smash{{\sP\kern-2.0pt}_2\kern-0.5pt(5)}}
\newcommand{\Ptwof}{\smash{{\sP\kern-2.0pt}_2\kern-0.5pt(6)}}
\newcommand{\Ptwokm}{\smash{{\sP\kern-2.0pt}_2\kern-0.5pt(2k-1)}}
\newcommand{\Pone}{\smash{{\sP\kern-2.5pt}_2\kern-0.5pt(n{-}1)}}

\newcommand{\Gr}{\smash{{\sG\kern-1.5pt}_q\kern-0.5pt(n,k)}}
\newcommand{\Gi}{\smash{{\sG\kern-1.5pt}_q\kern-0.5pt(n,i)}}
\newcommand{\Gj}{\smash{{\sG\kern-1.5pt}_q\kern-0.5pt(n,j)}}
\newcommand{\Grmk}{\smash{{\sG\kern-1.5pt}_q\kern-0.5pt(n,n-k)}}
\newcommand{\Grdk}{\smash{{\sG\kern-1.5pt}_q\kern-0.5pt(2k,k)}}
\newcommand{\Grekappa}{\smash{{\sG\kern-1.5pt}_q\kern-0.5pt(n,e+1-\kappa)}}
\newcommand{\Grtwoekappa}{\smash{{\sG\kern-1.5pt}_q\kern-0.5pt(n,2e+1-\kappa)}}
\newcommand{\Gremkappa}{\smash{{\sG\kern-1.5pt}_q\kern-0.5pt(n,e-\kappa)}}
\newcommand{\Gn}{\smash{{\sG\kern-1.5pt}_2\kern-0.5pt(n,n{-}1)}}
\newcommand{\Gnq}{\smash{{\sG\kern-1.5pt}_q\kern-0.5pt(n,n{-}1)}}
\newcommand{\Gone}{\smash{{\sG\kern-1.5pt}_2\kern-0.5pt(n,1)}}
\newcommand{\Gqone}{\smash{{\sG\kern-1.5pt}_q\kern-0.5pt(n,1)}}
\newcommand{\GTwo}{\smash{{\sG\kern-1.5pt}_2\kern-0.5pt(n,k)}}
\newcommand{\GTwonk}[2]{{\smash{{\sG\kern-1.5pt}_2\kern-0.5pt({#1},{#2})}}}
\newcommand{\Gnk}{\smash{{\sG\kern-1.5pt}_2\kern-0.5pt(n,n{-}k)}}
\newcommand{\Greone}{\smash{{\sG\kern-1.5pt}_q\kern-0.5pt(n,e{+}1)}}
\newcommand{\Gretwo}{\smash{{\sG\kern-1.5pt}_q\kern-0.5pt(n,e{+}2)}}

\newcommand{\be}[1]{\begin{equation}\label{#1}}
\newcommand{\ee}{\end{equation}}

\newcommand{\Cref}[1]{Co\-rol\-la\-ry\,\ref{#1}}

\newtheorem{theorem}{Theorem}
\newtheorem{lemma}{Lemma}

\newtheorem{corollary}{Corollary}

\newtheorem{definition}{Definition}

\begin{document}

\title{On Levenshtein Balls with Radius One}

\author{\IEEEauthorblockN{Daniella Bar-Lev}
\IEEEauthorblockA{Dept. of Computer Science\\
Technion-Israel Institute of Technology\\
Haifa 3200003, Israel \\
Email: daniellalev@cs.technion.ac.il} \and
\IEEEauthorblockN{Tuvi Etzion}
\IEEEauthorblockA{Dept. of Computer Science\\
Technion-Israel Institute of Technology\\
Haifa 3200003, Israel \\
Email: etzion@cs.technion.ac.il}  \and
\IEEEauthorblockN{Eitan Yaakobi}
\IEEEauthorblockA{Dept. of Computer Science\\
Technion-Israel Institute of Technology\\
Haifa 3200003, Israel \\
Email: yaakobi@cs.technion.ac.il}}

\maketitle
\begin{abstract}
The rapid development of DNA storage has brought the deletion and insertion channel, once again, to the front
line of research. When the number of deletions is equal to the number of insertions, the \emph{Fixed Length Levenshtein} (\emph{FLL})
metric is the right measure for the distance between two words of the same length. The size of a ball
is one of the most fundamental parameters in any metric. The size of the ball with radius one in the FLL metric depends on the number of runs and the length of the alternating segments of the given word. In this work, we find the minimum, maximum, and average size
of a ball with radius one, in the FLL metric. The related minimum and maximum sizes
of a maximal anticode with diameter one are also calculated.
\end{abstract}


\section{Introduction}

Coding for DNA storage has attracted significant attention in the previous decade due to recent experiments and demonstrations of the viability of storing information in macromolecules~\cite{CGK12,Getal13,Oetal17,YGM17,TWAEHLSZM19}. Given the trends in cost decreases of DNA synthesis and sequencing, it is estimated that already within this decade DNA storage may become a highly competitive archiving technology. However, DNA molecules induce error patterns that are fundamentally different from their digital counterparts~\cite{HMG18,SOSAYY19}. This distinction results from the specific error behavior in DNA and it is well known that errors in DNA are typically in the form of substitutions, insertions, and deletions, where most published studies report that deletions are the most prominent ones, depending upon the specific technology for synthesis and sequencing. Hence, coding for insertion and deletion errors has received renewed interest recently due to its high relevance to the error model in DNA storage; see e.g.~\cite{CS19,GW17,SB19,TPFV19}. This paper takes one more step in advancing this study and its goal is to study the size of the ball in the Fixed Length Levenshtein metric. 

If a word $\bfx \in \Z_m^n$ can be transferred to a word $\bfy \in \Z_m^n$ using $t$ deletions and $t$ insertions
(and cannot be transferred using a smaller number of deletions and insertions), then their
{\bf \emph{Fixed Length Levenshtein (FLL) distance}} is $t$, which is denoted by $d_\ell (\bfx,\bfy) =t$. 
Let $G=(V,E)$ be a graph whose set of vertices $V = \Z_m^n$
and two vertices $\bfx,\bfy \in V$ are connected by an edge if $d_\ell (\bfx,\bfy)=1$. The FLL
distance defines a \emph{graphic metric}, i.e., it is a metric and for each $\bfx,\bfy \in \Z_m^n$, $d_\ell (\bfx,\bfy)=t$
if and only if the length of the shortest path between $\bfx$ and $\bfy$ in $G$ is $t$.

One of the most fundamental parameters in any metric is the size of a ball with a given radius $t$ centered at a word~$\bfx$.
There are many metrics, e.g. the Hamming metric, the Johnson metric, or the Lee metric, where
the size of a ball does not depend on the word~$\bfx$. This is not the case in the FLL metric.
Moreover, the graph $G$ has a complex structure and it makes it much more difficult to find the
exact size of any ball and in particular the size of
a ball with minimum or maximum size. In~\cite{SaDo13}, a formula for the size of the ball with radius one in the FLL metric was given. This formula depends on the number of runs in the word and the lengths of its alternating segments. Nevertheless, it is still difficult to compute from this formula what the size of the maximum ball is. In this paper we find explicit expressions for the minimum and maximum sizes of a ball when the ball is of radius one. We also find the average size of a ball when the radius of the ball is one. Finally, we consider the related concept of anticode in the FLL distance and
find the maximum size and the minimum size of maximal anticodes with diameter one.

The rest of this paper is organized as follows. Section~\ref{sec:gen_comb} introduces some basic concepts
and observations required for our exposition. Section~\ref{sec:cods_and_balls} presents basic results and equivalence of codes, for
codes correcting deletions and insertions. The minimum size of a ball is discussed in Section~\ref{sec:min_size}, while
in Section~\ref{sec:anticodes} the minimum size and maximum size of maximal anticodes with diameter one are computed.
The maximum size and average size of balls with radius one are computed in Section~\ref{sec:bounds}. 

\section{Preliminaries}
\label{sec:gen_comb}

In this section we present the definitions and notations as well as several results that will be used throughout the paper.

Let $\Z_m$ denote the set of integers $\{0,1,\ldots,m-1\}$ and for an integer $n\ge0$,
let $\Z_m^n$ be the set of all sequences (words) of length $n$ over the alphabet $\Z_m$.
For an integer $t$, $0\le t\le n$, a sequence $\bfy\in\Z_m^{n-t}$ is a \emph{$t$-subsequence} of
$\bfx\in\Z_m^n$ if $\bfy$ can be obtained from $\bfx$ by deleting $t$ symbols from $\bfx$.
That is, there exist $n-t$ indices ${1\le i_1<i_2<\cdots<i_{n-t}\le n}$ such that $y_j=x_{i_j}$, for all $1\le j\le n-t$. We say that~$\bfy$ is a \emph{subsequence} of $\bfx$ if~$\bfy$ is a $t$-subsequence of $\bfx$ for some~$t$.
Similarly, a sequence $\bfy\in\Z_m^{n+t}$ is a \emph{$t$-supersequence} of~${\bfx\in\Z_m^n}$ if $\bfx$ is a $t$-subsequence of~$\bfy$.

For a sequence $\bfx\in\Z_m^n$, let $\bfx_{[i,j]}$ be the subsequence $x_ix_{i+1}\cdots x_{j}$ and for a set of indices $I\subseteq \{1,\ldots,n\}$, the sequence ${\bfx_|}_{I}$ is the \emph{projection} of $\bfx$ on the ordered indices of $I$, which is the subsequence of $\bfx$ received by the symbols in the entries of $I$. For a symbol ${\sigma\in \Z_m}$, $\sigma^n$ denotes the sequence with $n$ consecutive $\sigma$'s.

The \emph{Hamming weight} of a word $\bfx\in\Z_m^n$ is denoted by $\wt{\bfx}$ and is equal to the number of nonzero coordinates in~$\bfx$. The \emph{Hamming distance} between two words ${\bfx,\bfy\in\Z_m^n}$, denoted by $d_H(\bfx,\bfy)$, is the number of coordinates in which $\bfx$ and $\bfy$ differ.

\begin{definition}
The {\emph{\bf Hamming $\bft$-ball}} centered at ${\bfx\in\Z_m^n}$, $B_t(\bfx)$, is defined by
$$B_t(\bfx)\triangleq \left\{\bfy\in\Z_m^n\ :\  d_H(\bfx,\bfy)\le t \right\}.$$
\end{definition}

For $\bfx\in\Z_m^n$, the number of words in each Hamming $t$-ball is a function of only $n, m$ and $t$. The number of such words is
\begin{align}
\label{eq: hamming ball size}
 |B_t(\bfx)|=\sum_{i=0}^t\binom{n}{i}(m-1)^i.
\end{align}

\begin{definition}
The {\emph{\bf deletion $\bft$-sphere}} centered at ${\bfx\in\Z_m^n}$, $D_t(\bfx)\subseteq \Z_m^{n-t}$, is the set of all $t$-subsequences of~$\bfx$.
The {\emph{\bf  insertion $\bft$-sphere}} centered at ${\bfx\in\Z_m^n}$, $I_t(\bfx)\subseteq \Z_m^{n+t}$, is the set of all $t$-supersequences of $\bfx$.
\end{definition}

The following lemma was proven in~\cite{L01} and will be used in some of the proofs in the paper.

\begin{lemma}~\label{lem: del/ins intersection}
If $\bfx,\bfy \in \Z_2^n$ are distinct words, then
$$|D_1(\bfx)\cap D_1(\bfy)|\le 2\ \text{ and }\  |I_1(\bfx)\cap I_1(\bfy)|\le 2.$$
\end{lemma}

Note that for two sequences $\bfx,\bfy\in\Z_m^n$, the FLL distance between $\bfx$ and $\bfy$,
$d_\ell(\bfx,\bfy)$, is the smallest $t$ for which there exists a $t$-subsequence $\bfz\in\Z_m^{n-t}$  of both $\bfx$ and $\bfy$, i.e.,
\begin{equation}~\label{eq: deletion intersection}
d_{\ell}(\bfx,\bfy)= \min\{t': D_{t'}(\bfx)\cap D_{t'}(\bfy) \ne \varnothing\}.
\end{equation}
In other words, $t$ is the smallest integer such that it is possible to receive $\bfy$ from $\bfx$ by $t$ deletions and $t$ insertions. 

A \emph{longest common subsequence }(LCS) of sequences $\bfx_1,\ldots,\bfx_p\in \Z_m^n$ is a subsequence $\bfy$ of $\bfx_i$ for each ${1\le i\le p}$, where there is no other such subsequence $\bfz$ of a longer length.
In other words $\bfy \in \bigcap_{i=1}^p D_t(x_i)$ for some $t$ and $t$~is the smallest integer such that  ${\bigcap_{i=1}^p D_t(x_i)\ne \varnothing}$.
The set of LCSs of $\bfx_1,\ldots,\bfx_p$ is denoted by $\mathsf{LCS}( \bfx_1,\ldots,\bfx_p )$ and the length of any such LCS is denoted by $\ell\ell cs(\bfx_1,\ldots,\bfx_p)$. The following lemma is well known and can be easily verified.
\begin{lemma}\label{lem: deletion intersection and LCS}
For $\bfx,\bfy\in\Z_m^n$, $D_t(\bfx)\cap D_t(\bfy)=\varnothing$ if and only if ${\ell\ell cs(\bfx,\bfy)< n-t}$.
\end{lemma}

Combining (\ref{eq: deletion intersection}) and Lemma~\ref{lem: deletion intersection and LCS} implies the following known result.
\begin{corollary}~\label{cor: LCS length}
For all ${\bfx,\bfy \in\Z_m^n}$ we have that $$\ell\ell cs(\bfx,\bfy)= n-d_\ell(\bfx,\bfy).$$
\end{corollary}

For two sequences $\bfx\in \Z_m^{n_1}$ and $\bfy\in \Z_m^{n_2}$, $\ell\ell cs(\bfx,\bfy)$  is given by the following recursive formula

\vspace{-2ex}
\begin{footnotesize}
\begin{align}\label{eq: recursive LCS}
  &\ell\ell cs(\bfx,\bfy)= \nonumber \\
 &\begin{cases}
0 & n_1 = 0 \text{ or } n_2 = 0 \\
1 +\ell\ell cs( \bfx_{[1:{n_1}-1]}, \bfy_{[1:n_2-1]}) & x_{n_1}=y_{n_2}\\
\max \left\{
\ell\ell cs(\bfx_{[1:n_1-1]}, \bfy), \ell\ell cs(\bfx, \bfy_{[1:n_2-1]})
\right\} & \text{otherwise}
\end{cases}
\end{align}
\end{footnotesize}
\begin{definition}
The {\bf $ (\bft_1,\bft_2)$-deletion-insertion sphere} centered at $\bfx\in\Z_m^n$, $DI_{t_1,t_2}(\bfx)\subseteq \Z_m^{n-t_1+t_2}$, is the set of all the sequences that can be obtained from $\bfx$ by $t_1$ deletions and $t_2$ insertions.   
\end{definition}
\begin{definition}
The {\emph{\bf FLL $\bft$-ball}} centered at ${\bfx\in\Z_m^n}$, ${L_t(\bfx)\subseteq \Z_m^{n}}$, is defined by
$$L_t(\bfx) \triangleq \{  \bfy\in\Z_m^n \ : \  d_\ell(\bfx,\bfy)\leq t \} .$$
\end{definition}

For a sequence $\bfx\in\Z_m^n$,
a \emph{run} of $\bfx$ is a maximal subsequence $\bfx_{[i,j]}$ of identical symbols. The number of runs in $\bfx$ will be denoted by $\rho(\bfx)$.
We say that a subsequence $\bfx_{[i,j]}$ is an \emph{alternating segment}
if $\bfx_{[i,j]}$ is a sequence of alternating distinct symbols $\sigma,\sigma'\in \Z_m$.
Note that $\bfx_{[i,j]}$ is a maximal alternating segment if $\bfx_{[i,j]}$ is an alternating segment
and $\bfx_{[i-1,j]},\bfx_{[i,j+1]}$ are not. The number of maximal alternating segments of a sequence $\bfx$ will be denoted by $a(\bfx)$. For example, in $\bfx=00110100$, $\rho(\bfx) = 5$ and $a(\bfx) = 4$,  where the four maximal alternating segments are $0,01,1010,0$. Note that for binary sequences, $a(\bfx)  + \rho(\bfx) = |\bfx| +1$.  The next lemma states the known result from~\cite{SaDo13} on the size of the FLL $1$-ball.

\begin{lemma}{\cite{SaDo13}.}~\label{lem: L1 size}
 For all $\bfx\in\Z_m^n$, 
$$|L_1(\bfx)| = \rho(\bfx)\cdot (n(m-1)-1) + 2 - \sum_{i=1}^{a(\bfx)} \frac{(s_i-1)(s_i-2)}{2},$$
where $s_i$, $1\le i\le a(\bfx)$,
denotes the length of the i-th maximal alternating segment of $\bfx$.
\end{lemma}

An \emph{anticode of diameter $t$} in $\Z_m^n$ is a subset $\cA\subseteq \Z_m^n$
such that for any $\bfx,\bfx'\in \cA$, $d_\ell(\bfx,\bfx')\le t$.
We say that $\cA$ is a \emph{maximal anticode} if
there is no other anticode of diameter $t$ in  $\Z_m^n$ which contains $\cA$.

\section{Deletions/Insertions and the FLL Distance}
\label{sec:cods_and_balls}

A subset $\cC\subseteq\Sigma_q^n$ is a \emph{$t$-deletion-correcting code} (\emph{${t\text{-insertion-correcting code}}$}, respectively) if for any two distinct codewords $\bfc,\bfc'\in\cC$ we have that $D_t(\bfc)\cap D_t(\bfc')=\varnothing$ (${I_t(\bfc)\cap I_t(\bfc')=\varnothing}$, respectively).
Similarly, $\cC$ is called a \emph{$(t_1,t_2)$-deletion-insertion-correcting code} if for any two distinct codewords $\bfc,\bfc'\in\cC$ we have that 
$DI_{t_1,t_2}(\bfc)\cap DI_{t_1,t_2}(\bfc')=\varnothing$.
Levenshtein~\cite{L66} proved that $\cC$ is a $t$-deletion-correcting code if and only if $\cC$ is a $t$-insertion-correcting code and if and only if $\cC$ is a $(t_1,t_2)$-deletion-insertion-correcting code for every $t_1,t_2$ such that $t_1+t_2\le t$.
A straightforward generalization is the following result~\cite{CK13}.

\begin{lemma}
\label{lem: equivalent codes}
For all $t_1, t_2\in\Z$, if $\cC\subseteq\Z_m^n$ is a ${(t_1,t_2)\text{-deletion-insertion-correcting code}}$, then $\cC$ is also a $(t_1+t_2)$-deletion-correcting code.
\end{lemma}

\begin{corollary}
For $\mathcal{C}\subseteq \Z_m^n$, the following statements are equivalent.
\begin{enumerate}
\item $\cC$ is a $(t_1,t_2)$-deletion-insertion-correcting code. 
\item $\cC$ is a $(t_1+t_2)$-deletion-correcting code.
\item $\cC$ is a $(t_1+t_2)$-insertion-correcting code.
\item $\cC$ $(t_1',t_2')$-deletion-insertion-correcting code for any $t_1',t_2'$ such that $t_1'+t_2' = t_1+t_2$.
\end{enumerate}
\end{corollary}

\begin{lemma}
A code $\cC\in\Z_m^n$ is a $(2t+1)$-deletion-correcting code if and only if \\
$~~~\bullet$  $\cC$ is a $(t,t)$-deletion-insertion-correcting code \\
and also\\ 
$~~~\bullet$ if exactly $t+1$ FLL errors (i.e., $t+1$ insertions and $t+1$ deletions) occurred, then $\cC$ can detect these $t+1$ FLL errors.
\end{lemma}

\begin{IEEEproof}
If $\mathcal{C}$ is a $(2t+1)$-deletion-correcting code, then by definition for any $\bfc_1,\bfc_2\in \mathcal{C}$ we have that
$$
D_{2t+1}(\bfc_1)\cap D_{2t+1}(\bfc_2)=\varnothing.
$$
Therefore, by Lemma~\ref{lem: deletion intersection and LCS} for any two distinct codewords $\bfc_1, \bfc_2\in \mathcal{C}$ we have that $\ell \ell cs(\bfc_1,\bfc_2)\le n-(2t+1).$ Hence,  by Corollary~\ref{cor: LCS length}, ${d_\ell(\bfc_1,\bfc_2)\ge 2(t+1)}$.  Since the FLL metric is graphic, it follows that  $\mathcal{C}$ can correct up to $t$ FLL errors and if exactly $t+1$ FLL errors occurred it can detect them. 

For the other direction, assume that $\mathcal{C}$  is a $(t,t)$-deletion-insertion-correcting code and if exactly $t+1$ FLL errors occurred, then $\cC$ can detect them. By Lemma~\ref{lem: equivalent codes}, $\mathcal{C}$ 
is a $2t$-deletion-correcting code which implies that ${D_{2t}(\bfc_1)\cap D_{2t}(\bfc_2) = \varnothing}$ for all $\bfc_1,\bfc_2\in\cC$, and hence by~(\ref{eq: deletion intersection}) we have that
 $$\forall \bfc_1,\bfc_2\in \mathcal{C}: \ \ \ d_\ell(\bfc_1,\bfc_2) > 2t.$$
Let us assume to the contrary that there are two codewords  $\bfc_1,\bfc_2\in \cC$ such that $d_\ell(\bfc_1,\bfc_2)=2t+1$. Since the FLL metric is graphic, it follows that there exists a word $\bfy\in\Z_m^n$ such that $d_\ell(\bfc_1,\bfy) = t$ and $d_\ell(\bfy,\bfc_2)= t+1$  which contradicts the fact that up to~$t$ FLL errors can be corrected and exactly $t+1$ FLL errors can be detected.
Hence,
$$\forall \bfc_1,\bfc_2\in \cC: \ \ \ d_\ell(\bfc_1,\bfc_2) > 2t+1,$$
and by definition, $\mathcal{C}$ can correct $2t+1$ deletions.
\end{IEEEproof}

\section{Minimum Size of a Ball}
\label{sec:min_size}

The minimum size of the FLL $t$-ball is derived from the size of the Hamming ball.
Changing the symbol in the $i$-th position from $\sigma$ to $\sigma'$ in any sequence can be done by first deleting $\sigma$  and then inserting $\sigma'$. Hence we have the following lemma. 
\begin{lemma}
\label{lem: B_t <= L_t}
If $n\ge t\ge0$ are integers and $\bfx\in\Z_m^n$, then $B_t(\bfx)\subseteq L_t(\bfx)$ and hence  $|B_t(\bfx)|\le |L_t(\bfx)|$.
\end{lemma}

\begin{lemma}
If $n>t\ge0$ are integers, then $B_t(\bfx) = L_t(\bfx)$ if and only if $\bfx=\sigma^n$ for $\sigma\in\Z_m$.
\end{lemma}
\begin{IEEEproof}
Assume w.l.o.g that $\bfx=0^n$ and let $\bfy\in L_t(\bfx)$ be a sequence obtained from $\bfx$ by at most $t$ insertions and $t$ deletions. Hence, $ \wt \bfy \le t$ and $\bfy\in B_t(\bfx)$.
Therefore, Lemma~\ref{lem: B_t <= L_t} implies that $B_t(\bfx) = L_t(\bfx)$.

For the other direction, let $\bfx\in\Z_m^n$ were ${\bfx\ne \sigma^n}$ for all~$\sigma\in\Z_m$.
Since by Lemma~\ref{lem: B_t <= L_t}, $B_t(\bfx)\subseteq L_t(\bfx)$, to complete the proof, it is sufficient to show that there exists a sequence $\bfy\in L_t(\bfx)$\textbackslash $B_t(\bfx)$.
Denote $\bfx=(x_1,x_2,\ldots,x_n)$ and let $i$ be the smallest index for which $x_i\ne x_{i+1}$. Let $\bfy$ be the sequence defined by
$$ \bfy \triangleq \left(y_1,y_2,\ldots,y_{i-1},x_{i+1},x_{i}, y_{i+2},\ldots,y_{n}\right),$$
where $y_j\ne x_j$ for the first $t-1$ indices (for which ${j\notin\{ i,i+1\}}$) and $y_j=x_j$ otherwise.
Clearly, $\bfy$  differs from~$\bfx$ in $t+1$ indices and therefore $\bfy\notin B_t(\bfx)$.
On the other hand, $\bfy$ can be obtained from $\bfx$ by deleting $x_i$ and inserting it to the right of $x_{i+1}$ and then applying $t-1$ deletions and $t-1$ insertions whenever $y_j\ne x_j$ (for $j\notin\{i,i+1\}$). Thus, $\bfy\in  L_t(\bfx)$\textbackslash $B_t(\bfx)$ which completes the proof.
\end{IEEEproof}

\begin{corollary}
If $n>t\ge 0$ and $m>1$ are integers, then the size of the minimum FLL $t$-ball is
$$\min_{\bfx\in\Z_m^n}\left|L_t(\bfx)\right| = \sum_{i=0}^t\binom{n}{i}(m-1)^i,$$
and the minimum is obtained only by the balls centered at $\bfx=\sigma^n$ for any $\sigma\in\Z_m$.
\end{corollary}

\section{Binary Anticodes with Diameter one}
\label{sec:anticodes}

In this section we present tight lower and upper bounds on the size of maximal binary anticodes of diameter one in the FLL metric.
To prove these bounds we need some useful properties of anticodes with diameter one in the FLL metric.
\begin{lemma}
\label{lem: suffix 00}
If an anticode $\cA$ of diameter one contains three distinct words with the suffix 00 then there is at most one word in $\cA$ with the suffix 01.
\end{lemma}
\begin{IEEEproof}
Let $\bfa,\bfa',\bfa''\in\cA$ be three words with the suffix~{00}. Assume to the contrary that there exist two distinct words $\bfb,\bfb'\in \cA$ with the suffix {01}. Let $\bfy \in \mathsf{LCS}(\bfa,\bfb)$. By Corollary~\ref{cor: LCS length} the length of $\bfy$ is $n-1$ and since $\bfa$ ends with~{00}, $\bfy$ must end with {0} which implies that $\bfy=\bfb_{[1,n-1]}$. By the same arguments $\bfy\in \mathsf{LCS}(\bfb,\bfa')$ and $\bfy\in \mathsf{LCS}(\bfb, \bfa'')$.
Similarly,
$$\bfy' = \bfb'_{[1,n-1]}\in \mathsf{LCS}(\bfb',\bfa,\bfa',\bfa'').$$
Hence, $\bfa,\bfa',\bfa''\in I_1(\bfy) \cap I_1(\bfy') $ which is a contradiction to Lemma~\ref{lem: del/ins intersection} since $\bfy\ne \bfy'$. Thus, $\cA$ contains at most one word with the suffix {01}.
\end{IEEEproof}

\begin{lemma}~\label{lem: suffix 01}
If an anticode $\cA$ of diameter one contains three distinct words with the suffix {01}, then there is at most one word in $\cA$ with the suffix {00}.
\end{lemma}
\begin{IEEEproof}
Let $\bfa,\bfa',\bfa''\in\cA$ be three words with the suffix~{01}.
Assume to the contrary that there exist two distinct words $\bfb,\bfb'\in \cA$ with the suffix {00}.
For $\bfy\in \mathsf{LCS}(\bfa,\bfb)$, by Corollary~\ref{cor: LCS length} the length of $\bfy$ is $n-1$ and since $\bfb$ ends with~{00}, $\bfy$ must end with {0} which implies that $\bfy=\bfa_{[1,n-1]}$.
By the same arguments $\bfy\in \mathsf{LCS}(\bfa,\bfb')$.
Similarly,
\begin{align*}
\bfy' = \bfa'_{[1,n-1]} &\in \mathsf{LCS}(\bfa',\bfb,\bfb')\\
\bfy'' = \bfa''_{[1,n-1]} &\in \mathsf{LCS}(\bfa'',\bfb,\bfb').
\end{align*}
 Hence, $\bfy, \bfy', \bfy''\in D_1(\bfb) \cap D_1(\bfb') $ which is a contradiction to Lemma~\ref{lem: del/ins intersection}. Thus, $\cA$ contains at most one word with the suffix {00}.
\end{IEEEproof}

\begin{lemma}~\label{lem: suffixes}
Let $\cA$ be an anticode of diameter one. If ${\bfa,\bfa'\in \cA}$ are two distinct words that end with {00} and $\bfb,\bfb'\in\ \cA$ are two distinct words that end with {01},
then $\bfa_{[1,n-1]}\ne \bfb_{[1,n-1]}$ or  $\bfa'_{[1,n-1]}\ne\bfb'_{[1,n-1]}$.
\end{lemma}
\begin{IEEEproof}
Assume to the contrary that there exist $\bfa,\bfa',\bfb,\bfb'\in \cA$ such that  $\bfa_{[1,n-1]}=\bfb_{[1,n-1]}=\bfy0$ and  $\bfa'_{[1,n-1]}=\bfb'_{[1,n-1]}=\bfy'0$,
$\bfa,\bfa'$ end with {00}, and $\bfb,\bfb'$ end with {01}. Let,
\begin{align*}
\bfa \  &= a_1\ a_2 \cdots a_{n-2}\  {0\  0}   = \bfy\ \   {0\ 0} \\
\bfa'&= a'_1\ a'_2 \cdots a'_{n-2}\  {0\  0}=\bfy'\  {0\ 0} \\
\bfb \ &= a_1\ a_2 \cdots a_{n-2}\  {0\  1}  =\bfy\ \ {0\ 1} \\
\bfb'&= a'_1\ a'_2 \cdots a'_{n-2}\  {0\ 1}=\bfy'\  {0\ 1}.
\end{align*}
Notice that since the FLL distance between any two words in $\cA$ is one,
the Hamming weight of any two words can differ by at most one, which implies that $ \wt \bfy= \wt {\bfy'}$
(by considering the pairs $\bfa,\bfb'$ and $\bfa',\bfb$).
Clearly, $\bfy{0}\in \mathsf{LCS}(\bfa',\bfb)$ which implies that $\bfa'$ can be obtained from $\bfb$ by deleting the last~{1} of $\bfb$ and then inserting {0} into the LCS. Hence, there exists an index $0\le j\le n-2$ such that

\vspace{-2ex}
\begin{small}
\begin{equation}
\label{eq: a' from a}
a_1 a_2 \cdots a_j{0} a_{j+1} \cdots a_{n-2}{0}  = a_1'a_2'\cdots a'_j  a'_{j+1} \cdots a'_{n-2}{00}
\end{equation}
\end{small}

\vspace{-2ex}\noindent
Similarly, $\bfa$ can be obtained from $\bfb'$, i.e., there exists an index $0\le i\le n-2$ such that

\vspace{-2ex}
\begin{small}
\begin{equation}~\label{eq: a from a'}
a_1'a_2'\ldots a'_i {0} a'_{i+1} \cdots a'_{n-2}{0}  = a_1 a_2 \ldots a_i a_{i+1} \cdots a_{n-2} {00}.
\end{equation}
\end{small}

\vspace{-2ex}\noindent
Assume w.l.o.g. that $i\le j$. (\ref{eq: a' from a}) implies that $a_{r}=a_{r'}$ for $1\le r\le j$. In addition, $a_{n-2}={0}$ by (\ref{eq: a' from a}) and $a_{n-2}'={0}$ by (\ref{eq: a from a'}). By assigning  $a_{n-2} = a_{n-2}' = {0}$ into (\ref{eq: a' from a}) and (\ref{eq: a from a'}) we obtain that $a_{n-3}=a_{n-3}'={0}$.  Repeating this process implies that $a_{r}=a_{r'}={0}$ for $j+1 \le r\le n-2$.
Thus, it must be that $\bfy=\bfy'$ which is a contradiction.
\end{IEEEproof}

\begin{definition}
For an anticode ${\cA}\subseteq\Z_2^n$, the  \emph{\bf puncturing of~$\cA$ in the $n$-th coordinate}, $\cA'$, is defined by
$$\cA' \triangleq \left\{ \bfa_{[1:n-1]}\ : \ {\bfa}\in\cA \right\}.$$
\end{definition}

The following lemma is a conclusion of (\ref{eq: recursive LCS}) and Corollary~\ref{cor: LCS length}.
\begin{lemma}~\label{lem: anticode with fix last symbol}
Let $\cA\subseteq\Z_2^n$ be an anticode of diameter one. If the last symbol in all the words in $\cA$ is the same symbol ${\sigma \in \Z_2^n}$, then $\cA'$ is an anticode of diameter one and ${|\cA'|=|\cA|}$.
\end{lemma}


\begin{lemma}~\label{lem: anticode with alt suffix}
Let $\cA$ be an anticode of diameter one. If the suffix of each word in $\cA$ is either  {01} or  {10}, then $\cA'$ is an anticode of diameter one and $|\cA'|=|\cA|$.
\end{lemma}

\begin{IEEEproof}
Let $\bfa,\bfb\in\cA$ be two different words and let $\bfy \in \mathsf{LCS}(\bfa_{[1:n-1]}, \bfb_{[1:n-1]})$.  By (\ref{eq: recursive LCS}), $\ell\ell cs(\bfa,\bfb)\le |\bfy|+1$ and since $d_\ell(\bfa,\bfb)=1$ it follows that $|\bfy|\ge n-2$ and that $d_\ell(\bfa_{[1:n-1]},\bfb_{[1:n-1]})\le 1$. Thus, $\cA'$ is an anticode of diameter one. It is readily verified that $|\cA'|=|\cA|$.
\end{IEEEproof}

\begin{theorem}
Let $n>1$ be an integer and let ${\cA\subseteq\Z_2^n}$ be a maximal anticode of diameter one. 
Then, $|\cA|\le n+1$, and there exists a maximal anticode with exactly $n+1$ codewords.
\end{theorem}

\begin{IEEEproof}
Since two words $\bfx,\bfy$ such that $\bfx$ ends with {00} and $\bfy$ ends with {11} are at FLL distance at least $2$, w.l.o.g. assume that $\cA$ does not contain codewords that end with~{11}. It is easy to verify that the theorem holds for $n\in\{2,3,4\}$. Assume that the theorem does not hold and let $n^*>4$ be the smallest integer such that there exists an anticode  $\cA\subseteq\Z_2^{n^*}$ such that $|\cA|= n^*+2$. Since there are only three possible options for the last two symbols of codewords in $\cA$, there exist three different codewords in $\cA$ with the same suffix of two symbols.   \\
\textbf{Case $\bf 1$ - } Assume $\bfx,\bfy,\bfz\in \cA$ are  three different words with the suffix {00}. By Lemma~\ref{lem: suffix 00}, there exists at most one codeword in $\cA$ with the suffix {01} and since $\cA$ does not contain codewords with the suffix {11}, there exists at most one codeword in $\cA$ that ends with the symbol {1}. That is, there are~$n^*+1$ codewords with {0} as the last symbol. Denote this set of $n^*+1$ codewords by $\cA_1$. As a subset of the anticode~$\cA$, $\cA_1$ is also an anticode and hence by Lemma~\ref{lem: anticode with fix last symbol},~$\cA_1'$ is an anticode of length $n^*-1$ and size $n^*+1$ which is a contradiction to the minimality of $n^*$.\\
\textbf{Case $\bf 2$ - } Assume $\bfx,\bfy,\bfz\in \cA$ are  three different words with the suffix {01}.
By Lemma~\ref{lem: suffix 01}, there exists at most one codeword in $\cA$ with the suffix {00} and since $\cA$ does not contain codewords with the suffix {11} there exist $n^*+1$ codewords that end with either {01} or {10}.  Denote this set of $n^*+1$ codewords as $\cA_1$. As a subset of the anticode $\cA$, $\cA_1$ is also an anticode and hence by  Lemma~\ref{lem: anticode with alt suffix}, $\cA_1'$ is an anticode of length $n^*-1$ and size $n^*+1$ which is a contradiction to the minimality of $n^*$.\\
\textbf{Case $\bf 3$ - } Assume $\bfx,\bfy,\bfz\in \cA$ are  three different words with the suffix {10}.
By the previous two cases, there exist at most two codewords in $\cA$ with the suffix {00} and at most two codewords with the suffix {01}. Since there are no codewords with the suffix {11}, it follows that the number of words that end with {1} is at most two.
If there exist at most one codeword in $\cA$ that ends with {1}, then there are $n^*+1$ codewords in~$\cA$ that end with {0} and as in the first case, this leads to a contradiction.
Otherwise there are exactly two codewords in~$\cA$ with the suffix {01}. If there are less than two codewords with the suffix {00}, then, the number of codewords with suffixes {01} and {10} is at least $n^*+1$ and similarly to the second case, this is a contradiction to the minimality of~$n^*$.
Hence, there exist exactly two codewords in $\cA$ with the suffix~{00}.
There are exactly $n^*-2$ codewords in $\cA$ with the suffix {10} and two more codewords with the suffix {01}.
By Lemma~\ref{lem: anticode with alt suffix} the words in $\cA'$ that were obtained from these $n^*$ codewords are all different and have FLL distance one from each other. In addition, by Lemma~\ref{lem: suffixes},
 the prefix of length $n^*-1$ of at least one of the codewords that end with~{00} is different from the prefixes of length $n^*-1$ of the codewords that end with {01}. This prefix also differs from the prefixes of the codewords that end with {10}.
Therefore, $\cA'$ is an anticode with $n^*+1$ different codewords which is a contradiction to the minimality of~$n^*$. 

Note that the set $\cA=\left\{\bfa\in\Sigma_2^n \ : \wt \bfa \le1\right\}$ is an  anticode of diameter one with exactly $n+1$ codewords. Thus, the maximum size of an anticode of diameter one is~$n+1$.
\end{IEEEproof}

\begin{theorem}
Let $n>2$ be a positive integer and let $\cA \subseteq\Z_2^n$ be a maximal anticode of diameter one.
Then, $|\cA|\ge 4$ and there exists a maximal anticode with exactly 4 codewords.
\end{theorem}

\section{Balls with Radius one}
\label{sec:bounds}

\subsection{Balls with Maximum Size}

The following theorem was proved in~\cite{SGD14}.
\begin{theorem}
If $n$ and $m>2$ are positive integers, then the maximal FLL $1$-balls
are the balls centered at ${\bfx\in\Z_m^n}$,
such that the number of runs in $\bfx$ is $n$ and ${x_i\ne x_{i+2}}$  for all $1\le i\le n-2$.
In addition, the size of such a maximum $1$-ball is,
$$\max_{\bfx\in\Z_m^n}\abs{L_1(\bfx)} = n^2(m-1) - n + 2.$$
\end{theorem}

For the binary case, there exists a ball whose size is larger than the ball considered in~\cite{SGD14}.
The analysis to find such a ball is slightly more difficult, since by definition of a run, there is no sequence $\bfx$ with $n$ runs such that $x_i\ne x_{i+2}$
for some $i$.
The following lemmas lead to the main theorem of this subsection.

\begin{definition}
For $t\in\mathbb{N}$,  $\bfx\in\Z_2^n$ is an \emph{\bf $\alpha$-balanced sequence} if ${a(\bfx)=\alpha}$
and ${s_i\in \{\lceil\frac{n}{\alpha}\rceil, \lceil\frac{n}{\alpha}\rceil-1\}}$ for all $i\in\{1,\ldots,\alpha\}$, where $s_i$ was defined in Lemma~\ref{lem: L1 size}.
\end{definition}

\begin{lemma}
\label{cla: q=2 max ball for fix k}
If $n$ and $1\le \alpha\le n$ are positive integers then
$$\argmax_{\substack{\bfx\in\Z_2^n \\ a(\bfx)=\alpha}}|L_1(\bfx)| = \left\{\bfx\in\Z_2^n:
\bfx\text{ is an } \alpha\text{-balanced sequence}  \right\}.$$
\end{lemma}

\begin{lemma}
\label{cla: k balanced ball size}
If $\bfx^{(\alpha)}$ is an $\alpha$-balanced sequence of length $n$ then
\begin{small}
\begin{align*}
|L_1(\bfx^{(\alpha)})| & = (n+1-\alpha)(n-1) +2
 - \frac{k}{2}\left(\left\lceil\frac{n}{\alpha}\right\rceil-1\right)
\left(\left\lceil\frac{n}{\alpha}\right\rceil-2\right) \\
 & - \frac{\alpha-k}{2}\left(\left\lceil\frac{n}{\alpha}\right\rceil-2\right)
 \left(\left\lceil\frac{n}{\alpha}\right\rceil-3\right),
 \end{align*}
 \end{small}
 
 \vspace{-2ex}
\noindent where
$k\equiv n\pmod \alpha$ and $1\le k\le \alpha$.
\end{lemma}

By Lemma~\ref{cla: q=2 max ball for fix k}
\begin{small}
\begin{align*}
    \max_{x\in\Z_2^n}|L_1(\bfx)| &
    = \max_{1\le \alpha\le n}\left\{ \max_{\substack{\bfx\in\Z_2^n \\ \alpha(\bfx)=\alpha}}|L_1(\bfx)|\right\}
    = \max_{1\le \alpha\le n}\left|L_1(\bfx^{(\alpha)})\right|,
\end{align*}
\end{small}
 
\vspace{-2ex}
\noindent and the size $|L_1(\bfx^{(\alpha)})|$  for $1\le \alpha\le n$ is given by Lemma~\ref{cla: k balanced ball size}.
Hence, our goal is to find the set
$$\mathsf{T}(n) \triangleq  \argmax_{1\le \alpha\le n}\left|L_1(\bfx^{(\alpha)})\right|.$$

\begin{lemma}
\label{cla:maxk}
If $\bfx^{(\alpha)}$ is an $\alpha$-balanced sequence of length $n$, then 
\[|L_1(\bfx^{(\alpha)})|> |L_1(\bfx^{(\alpha-1)})|\]
if and only if $n>2(t-1)t$.
\end{lemma}

\begin{theorem}
\label{the: q=2 max ball}
Let $n$ be an integer. It holds that 
$$\mathsf{T}(n) = \argmin_{\alpha\in\mathbb{N}}\left|\alpha-\frac{1}{2}\sqrt{1+2n}\right|.$$
Consequently, the maximum FLL $1$-balls are the balls centered at the $\alpha$-balanced sequences of length $n$,
for any $\alpha\in\mathsf{T}(n)$ and

\vspace{-2ex}
\begin{small}
\begin{align*}
& \max_{\bfx\in\Z_2^n}  |L_1(\bfx)|  = n^2 -n(\alpha+1) +\alpha+ 2 \\
& \ \  - \frac{k}{2}\left(\left\lceil\frac{n}{\alpha}\right\rceil-1\right)
\left(\left\lceil\frac{n}{\alpha}\right\rceil-2\right)
- \frac{\alpha-k}{2}\left(\left\lceil\frac{n}{\alpha}\right\rceil-2\right)
\left(\left\lceil\frac{n}{\alpha}\right\rceil-3\right),
\end{align*}
\end{small}

\vspace{-2ex}
\noindent where
$k\equiv n\pmod \alpha$ and $1\le k\le \alpha$.
\end{theorem}
\noindent 
{\bf{Note:}}  $\mathsf{T}(n)$ in Theorem~\ref{the: q=2 max ball} can be either of size one or of size two.

\begin{corollary} 
 If $n$ is a sufficiently large integer, then
$$\max_{x\in\Sigma_2^n}\left\{|L_1(x)|\right\} =  n^2 - \sqrt{2}n^{\frac{3}{2}}+O(n).$$
\end{corollary}

\subsection{The Average Size of a Ball}

By Lemma~\ref{lem: L1 size}, for any $\bfx\in\Z_m^n$
\begin{small}
\begin{align*}
	 \ |L_1(\bfx)| =  \rho(\bfx)(nm-n-1) + 2 -\frac{1}{2} \sum_{i=1}^{a(\bfx)} s_i^2 + \frac{3}{2} \sum_{i=1}^{a(\bfx)} s_i-  a(\bfx).
\end{align*}
\end{small}
Thus, the average size of the FLL $1$-ball is
\begin{small}
\begin{align*}
  \mathop{{}\mathbb{E}}_{\bfx\in\Z_m^n}\left[\rho(\bfx)(n(m-1)-1) + 2
	-\frac{1}{2} \sum_{i=1}^{a(\bfx)} s_i^2 + \frac{3}{2} \sum_{i=1}^{a(\bfx)} s_i-  a(\bfx)\right].
\end{align*}
\end{small}

The following lemma leads to our main theorem of this subsection.
\begin{lemma}\label{lem: avg vals}
For any integers $n,m>1$, we have that
\begin{align*}
{1)\ } & \mathop{{}\mathbb{E}}_{\bfx\in\Z_m^n}\left[\sum_{i=1}^{a(\bfx)}s_i\right] = n + (n-2)\cdot \frac{(m-1)(m-2)}{m^2},\\
{2)\ }& \mathop{{}\mathbb{E}}_{\bfx\in\Z_m^n}\left[a(\bfx)\right] =  1 +  \frac{(n-2)(m-1)(m-2)}{m^2} + \frac{n-1}{m},\\
{3)\ } & \mathop{{}\mathbb{E}}_{\bfx\in\Z_m^n}\left[\rho(\bfx)\right] = n - \frac{n-1}{m},\\
{4)\ } & \mathop{{}\mathbb{E}}_{\bfx\in\Z_m^n}\left[ \sum_{i=1}^{a(\bfx)}s_i^2\right] = \frac{n(4m^2-3m+2)}{m^2}\\ & \ \ \ \ \ \ \ \ \ \ \ \ \ \ \ \ \ \ \ +
   \frac{6m - 4}{m^2}
   - 4
   -\frac{2}{m - 1} \left( 1 - \frac{1}{m^n}\right).
\end{align*}
\end{lemma}

\begin{theorem}
For any integers $n,m>1$, we have that
\begin{small}
\begin{align*}
\mathop{{}\mathbb{E}}_{\bfx\in\Z_m^n}\left[|L_1(\bfx)|\right]
& =  n^2\left(m+\frac{1}{m} -2\right) - \frac{n}{m} - \frac{(m-1)(m-2)}{m^2}\\ &
	 +3 - \frac{3}{m} + \frac{2}{m^2} + \frac{m^n-1}{m^n(m-1)}.
	 \end{align*}
	 	 \end{small}
\end{theorem}


\section*{Acknowledgment}
D. Bar-Lev and T. Etzion were supported by ISF grant no. 222/19.
E. Yaakobi was supported by the United States-Israel BSF grant no. 2018048.

\end{document}